\documentclass[rmp,aps,twocolumn,nofootinbib,endfloats,showpacs]{revtex4}
\usepackage{amsmath}
\usepackage{textcomp}
\begin{document}

\title{An Introduction to Consistent Quantum Theory}

\author{P C Hohenberg}

\affiliation{Department of Physics, New York University, New York, NY}

\date{\today}

\begin{abstract}
This paper presents an elementary introduction to Consistent Quantum Theory (CQT), as developed by Griffiths and others over the past 25 years. The theory is a version of orthodox (Copenhagen) quantum mechanics, based on the notion that the unique and mysterious feature of quantum, as opposed to classical, systems is the simultaneous existence of multiple incompatible representations of reality, referred to as ``frameworks''. A framework is a maximal set of properties of a system for which probabilities can be consistently defined. This notion is expressed by saying that a framework provides an exhaustive set of exclusive alternatives (ESEA), but no single framework suffices to fully characterize a quantum system. Any prediction of the theory must be confined to a single framework and combining elements from different frameworks leads to quantum mechanically meaningless statements. This ``single framework rule'' is the precise mathematical statement of Bohr's complementarity. It is shown that if the microscopic description is assumed to incorporate these elements in a \emph{local} setting, then distant entanglements, macroscopic measurements, wave function collapse and other mysterious features of quantum behavior follow in a logical manner. The essential elements of the theory are first explained using the simplest quantum system, a single spin-\mbox{\textonehalf} degree of freedom at one time.

\end{abstract}

\pacs{03.65.Ta}
\maketitle
\section{\label{1}Introduction}
The present author considers Consistent Quantum Theory \cite{grif1} (hereafter CQT) to be the clearest and most correct formulation of non-relativistic quantum mechanics. The theory was introduced over 25 years ago by Robert Griffiths \cite{grif2} and further elaborated by \textcite{gh3}(see also \cite{jh1,jh2}) and by \textcite{o4}, but it appears to be little known outside of a rather narrow circle of specialists in foundations of quantum mechanics. Even within that community, however, CQT is mostly the object of rather cursory judgments and comments, rather than penetrating analysis and/or criticism. The theory is meant to be a version of orthodox quantum mechanics, formulated in such a way as to remove as much of the obscurity and mystery as possible (but no more!) and to state precisely what is assumed and what is derived. In the words of its inventor CQT is ``Copenhagen done right!''

This paper is an attempt to familiarize a broader community of students, teachers and researchers with the theory, in the hopes that they will find it as appealing as does the present author. The only original element that can be claimed for this presentation is the contention that the essence of quantum mechanics, including all its ``weirdness and mystery'', must be confronted and can be explained by studying the simplest example of a quantum mechanical system, a single spin-\mbox{\textonehalf} degree of freedom at one time. This is so because this system already displays the fundamental uniqueness of quantum mechanics, namely complementarity. Thus, questions of measurement, entanglement, nonlocality, collapse of the wave function and the myriad quantum paradoxes that have plagued the ``interpretation'' of the theory for over 80 years become add-ons that are relatively easily incorporated, once one has faced the fundamental issue of complementarity which already arises for the simplest system.

Our formulation of quantum mechanics is based on the following assumptions:
\begin{quote}
A.	Physical objects and their properties are represented by states in an appropriate (Hilbert) space.

\vspace{3 mm}
B.	The predictions of quantum mechanics are not deterministic but intrinsically probabilistic. They can (and should) be expressed as the probability that the system under study will possess a specific property at a given time, with suitable generalization for multiple times. This is so even though quantum mechanical states evolve according to the \emph{deterministic} Schroedinger equation, since the wave function yields probabilistic information about physical systems.

\vspace{3mm}
C.	The principle of unicity does not hold: there is not a unique exhaustive description of a physical system or process. Instead reality is such that it can be described in various alternative incompatible ways using descriptions, called ``frameworks'', which cannot be compared or combined. This is the principle of complementarity or incompatibility, expressed in terms of the ``single framework rule'', to be elucidated in what follows.
\end{quote}

As explained below, CQT takes a realistic point of view, in that it deals with real (i.e. intrinsic) properties of physical systems, independent of observers or measurement. The essential feature of the theory is that its ``quantum realism'' posits multiple incompatible but coexisting layers of reality, expressed in the theory as different ``frameworks''. Any quantum mechanical statement must refer to one and only one framework (the ``single framework rule''). This basic assumption is consistent with Bohr's complementarity principle, but it is formulated without recourse to measurement or to the existence of a classical regime and it is therefore more general and can be more precisely formulated than Bohr's complementarity principle.

The above, of course, assumes that quantum mechanics is an exact theory in the non-relativistic domain. Various authors have explored the possibility of physical corrections to quantum mechanics, which would in particular bring about collapse of the wave function when macroscopic measurements are made (see for example \textcite{ab5} and references therein), but we shall not consider this alternative in the present discussion.

\section {\label{2} A spin-\mbox{\textonehalf} particle}
The classical representation of a spin of magnitude \mbox{\textonehalf} can be taken as a vector in 3-space with magnitude \mbox{\textonehalf}, i.e. a point on the surface of a sphere of radius \mbox{\textonehalf}. The direction $\bf{w}$, for example, leads to the spin
\begin{equation}\label{e1}
 {\bf S_w}=[S_x{\bf x}+S_y{\bf y}+S_z{\bf z}],
\end{equation}
where
\begin{equation}\label{e2}
S_x^2+S_y^2+S_z^2=1/4,
\end{equation}
and $\bf{x, y, z}$  are unit vectors forming a basis in 3-space. Classical mechanics describes the motion (orbit) of this point on the surface of the sphere and a measurement, if carried out appropriately, would verify the predictions of that theory. A feature of the states in Eq.(\ref{e1}) is that for $\bf{S}$ in the z-direction, for example, we have
\begin{equation} \label{e3}
S_z=\mbox{\textonehalf}, \quad S_x=0,\quad S_y=0,
\end{equation}
i.e. the information regarding $S_x,S_y$ and $S_z$ is consistent and mutually compatible. A general state can have any value of $S_x,S_y$ and $S_z$   satisfying the constraint (\ref{e2}).

Quantum mechanically, the system is represented by states, also indexed by a direction $\bf{w}$ in 3-space, but these states exist in a discrete two-dimensional Hilbert space. Any direction $\bf{w}$ defines a basis $|w^+\rangle ,  |w^-\rangle$, such that any state of the system, corresponding to the direction $\bf{v}$, say, may be written
\begin{equation} \label{e4}
|\bf{v}\rangle=|v^+\rangle=\alpha(v,w)|w^+\rangle+\beta(v,w)|w^-\rangle,
\end{equation}
with
\begin{equation}\label{e5}
|\alpha|^2+|\beta|^2=1.
\end{equation}
The space of states possesses an inner product according to which the basis vectors satisfy the orthonormality conditions
\begin{subequations}
\begin{eqnarray}\label{e6}
\langle w^+|w^+\rangle&=&\langle w^-|w^-\rangle=1, \\
\langle w^+|w^-\rangle&=&\langle w^-|w^+\rangle=0.
\end{eqnarray}
\end{subequations}
The spin in direction $\bf{w}$, say, is an observable represented by an operator  $S_w$ with the property
\begin{equation}\label{e7}
S_w|w^+\rangle=\mbox{\textonehalf}|w^+\rangle,\quad S_w|w^-\rangle=-\mbox{\textonehalf}|w^-\rangle,
\end{equation}
which is expressed by saying that $|w^+\rangle$ and $|w^-\rangle$, are eigenstates of $S_w$ with eigenvalues $ \pm{\mbox{\textonehalf}}$ . The expectation value, or average value, $\langle S_w \rangle$ of  $S_w$ in an arbitrary state  $ |v\rangle $ is given by $\langle S_w \rangle = \langle v|S_w |v\rangle$, so it is $\pm{\mbox{\textonehalf}}$ in the states $|w^\pm\rangle$, respectively. This is quite different from the classical case described in Eqs.(\ref{e1}-\ref{e2}).

The basic assumption B stated in the Introduction is embodied in the Born Rule, which specifies that the probability that the operator $S_w$ has the value \mbox{\textonehalf} in the state $|\bf{v}\rangle$  is given by
\begin{equation}\label{e8}
\text{Prob}(S_w =\mbox{\textonehalf})=|\langle w^+|\bf{v}\rangle|^2=|\alpha|^2,
\end{equation}
where Eq.(\ref{e4}) has been used.Correspondingly the probability of the value -\mbox{\textonehalf} is $|\beta|^2$  and by Eq.(\ref{e5}) those are the only possibilities for $S_w$, since the probabilities add up to unity. In accordance with assumption B, probability is an intrinsic concept in the theory and not reducible to other notions.

Let us specify $v=z$ and $w=x$ or $y$. Then the commutation relations between spin components $S_x, S_y \;\text{and}\; S_z$ are realized by
\begin{subequations}
\begin{eqnarray}\label{e9}
|z^+\rangle=(1/\sqrt{2})(|x^+\rangle+|x^-\rangle)=(1/\sqrt{2})(|y^+\rangle-|y^-\rangle),\\
|z^-\rangle=(1/\sqrt{2})(|x^+\rangle-|x^-\rangle)=(1/\sqrt{2})(|y^+\rangle+|y^-\rangle),
\end{eqnarray}
\end{subequations}
If the system is in the state $|z^+\rangle$, then the following is true:
\begin{subequations}
\begin{eqnarray}\label{e10}
\text {Prob} (S_z=\mbox{\textonehalf})=1, \quad \text{Prob} (S_z = -\mbox{\textonehalf}) = 0,\\
\text {Prob} (S_x=\mbox{\textonehalf})=\mbox{\textonehalf}, \quad \text{Prob} (S_x  = -\mbox{\textonehalf}) = \mbox{\textonehalf},\\
\text {Prob} (S_y =\mbox{\textonehalf})=\mbox{\textonehalf},\quad \text{Prob} (S_y = -\mbox{\textonehalf}) = \mbox{\textonehalf}.
\end{eqnarray}
\end{subequations}
Note also that in contrast to Eq.(\ref{e2}), we have $\langle S^2 \rangle = 3/4$ for spin-\mbox{\textonehalf} in the quantum case.
\section{\label{3}The orthodox formulation of quantum mechanics}
According to the orthodox formulation of quantum mechanics (see, for example \textcite{ll6}), a system is ``represented'' by its wave function or state vectors which we denote by $|v\rangle$, say. The physical meaning of this state vector only becomes apparent when one makes a measurement, for example of $S_w$. Then the state of the system ``collapses'' to one of the \emph{eigenstates}  $|w^{\pm}\rangle$ of $S_w$, with probability given by the Born rule, and the result of the measurement is the corresponding eigenvalue. Thus, for example, if the state of the system is $|z^+\rangle$  and I measure $S_x$, then the system goes to states  $ |x^\pm\rangle$ with eigenvalues $\pm\mbox{\textonehalf}$, respectively, each with probability \mbox{\textonehalf}. The orthodox formulation has been criticized by many authors (see e.g. \textcite{grif1, gh3,o4,b7}) because the dynamical mechanism for the collapse remains undefined, and because of the heavy reliance on macroscopic measurements and on the classical domain in defining the physical meaning of the theory.

\section{\label{4}Complementarity}

An important distinction between the classical and quantum formulations described above lies in the question of compatibility between the values of $S_x, S_y$ and $\ S_z  $. In the classical case the three statements in Eq.(\ref{e3}), for example, are all compatible, whereas the corresponding quantum mechanical statements in Eqs.(10) cannot be combined into a single probabilistic formulation involving all three components of $\bf{S}$. The reason is that each line of Eq.(10) refers to a \emph{different sample space} and in contrast to classical mechanics these sample spaces cannot be consistently enlarged (``refined''), when the operators involved \emph{do not commute} (see below).

The above is an aspect of complementarity, which is the essence of quantum mechanics and it is a good idea to incorporate such a fundamental feature in the very formulation of the theory, rather than as an after-thought once one has gotten into trouble. Complementarity is at the heart of the Heisenberg Uncertainty (or better, Indeterminacy) Principle, according to which two observables represented by non-commuting operators (e.g. $S_x$ and $S_y$) not only cannot be \emph {measured} simultaneously with arbitrary precision, they cannot be \emph {defined} simultaneously with arbitrary precision. Note that we shall often use the term ``complementary'' to refer to the incompatibility of any two noncommuting variables, not just to variables such as $\bf{x}$ and $\bf{p}$ that are maximally incompatible.

\section{\label{5}Consistent Quantum Theory (CQT)}

How does one incorporate complementarity into the theory early on? A clue comes from the existence in the Hilbert space of an infinity of bases, each one of which can be used to represent any state of the system. For the simple case of a spin-\mbox{\textonehalf} system we define a ``framework'' as any basis in which we choose to represent a given state of the system. [This definition will be generalized in Sec. VIII below]. Since according to Eq.(\ref{e4}) an arbitrary quantum state $|v\rangle$ can be expanded in an infinite number of ways by choosing an appropriate basis $|w^{\pm}\rangle$, we see that the different frameworks are merely different ways of representing an arbitrary state in the Hilbert space of spin-\mbox{\textonehalf}. The characteristics of a framework in this simple case result from those of the basis, namely (i) that any state can be represented and (ii) the disjoint possibilities associated with that framework (spin up or spin down) have probabilities adding up to unity (certainty). Thus a framework provides an Exhaustive Set of Exclusive Alternatives (ESEA). It is exhaustive because of (i) and (ii), and exclusive because the basis vectors are mutually orthogonal. Note that although there are an infinite number of different frameworks in which to represent a given state, since each one is exhaustive any quantum mechanical statement only makes sense (and thus can only be correct) relative to a particular framework.
To illustrate the last statement let us associate with the state $|x^+\rangle$ the ``property'' $[x^+]$ which for the present discussion we define to be the projection operator onto the state $|x^+\rangle$. Then the properties $[x^+]$ and $[x^-]$ represent Boolean alternatives and we can apply the ordinary rules of logic to these operators. This logic refers to the x-framework. Alternatively, we can expand the state in question in the y-basis and employ the y-framework, but it is impossible to define a framework that combines the properties $[x^+]$ and $[y^+]$ in a consistent manner, since the projection operators do not commute. Indeed, any attempt to define the disjunction of properties $[x^+]\vee[y^+]$ (either $ S_x =\mbox{\textonehalf}\; \text{or} \; S_y =\mbox{\textonehalf}$) or the conjunction $[x^+]\wedge[y^+]$ ($ S_x =\mbox{\textonehalf}\; \text{and} \; S_y =\mbox{\textonehalf})$ will fail, since there is no (x,y)-framework or (x,y) basis in the Hilbert space.
In CQT the essential feature of quantum mechanics, indeed the full mystery and weirdness of the theory, is captured by the existence of \emph {multiple frameworks}, each one of which provides an account of quantum mechanical ``truth'', but different frameworks are mutually incompatible. A quantum mechanical statement is only meaningful \emph {relative} to a particular framework and it is meaningless (quantum mechanical nonsense) relative to a different framework. Thus asking ``what is the probability that the system has the property $[x^+]$?'' has no meaning in the y-framework. It is, however a perfectly meaningful question in the x-framework. Indeed, for a system prepared in the state $|z^+\rangle$ I may make the (x-framework) statement Prob($[x^+]$) = \mbox{\textonehalf} and the (y-framework) statement Prob($[y^+]$) = \mbox{\textonehalf}, but as noted above, I cannot give meaning to the probability of ``$[x^+]$ or $[y^+]$''.
Similarly, for a spin prepared in the state $|z^+\rangle$, the statements in Eq. (10) ``the system has the property $[z^+]$ with probability 1'' and ``the system has the property $[x^+]$ with probability \mbox{\textonehalf}'' are not contradictory but complementary, since they refer to different frameworks. This means that contrary to the classical case there exists no overall framework of which each line in (10) would be a component.

In concluding this section let us note that although each framework is associated with one and only one basis in the Hilbert space, a framework and a basis are not identical. Indeed, since the full wave function can be expanded in any basis the different bases are in some sense equivalent. A framework, on the other hand, identifies the properties (eigenstates) associated with the corresponding basis, but only retains the real probability for those properties and discards the remaining information. Thus different frameworks capture different real aspects of a quantum system and a full description of that system requires the set of all frameworks.

\section{\label{6}	Measurements}

As is well known any component of spin can be measured using a suitable Stern-Gerlach apparatus (see \textcite{grif1} or \textcite{ll6}). Notice, however, that the preceding discussion of CQT made no reference to measurements. This is because from the point of view of CQT measurements play no special role, just as in classical mechanics. This means that since measurements are just one type of physical interaction they must also be understood and interpreted within an appropriate framework. Although a proper description of such physical interactions requires the more general formalism of Sec.VIII below, since the system must now include a measurement apparatus, we can anticipate that a measurement of $S_x$, for example, needs to be interpreted in the x-framework. It will not be informative if analyzed in the z-framework.

As noted above, a measurement of $S_x$ is a physical process which causes its different eigenstates to be correlated with states of a macroscopic measurement apparatus. These correlations are not problematic when viewed in the appropriate framework. Rather than to say
\begin{quote}
 ``given that the system has been prepared in the state $|z^+\rangle$, I now make a measurement of $S_x$, which collapses the state $|z^+\rangle$ to $|x^+\rangle$  or  $|x^-\rangle$, each with probability  $\mbox{\textonehalf}$'',
\end{quote}
CQT says
\begin{quote}
``in the x-framework any state has two disjoint possibilities or properties $[x^+] \; \text{and} \; [x^-]$, so for a system prepared in the state $|z^+\rangle$ a  measurement of $S_x$, when viewed in the x-framework, will reveal one or the other of these properties, each with probability  $\mbox{\textonehalf}$. In the z-framework, on the other hand, I am unable to interpret an $S_x$ measurement''.
\end{quote}
These issues will be taken up again in Sec. IX, below, once the more general formulation of CQT has been presented.

\section{\label{7}	Einstein, Bohr and Bell}
The Einstein, Podolsky, Rosen (EPR) \cite{epr} argument starts from two basic assumptions:
\begin{quote}
{\bf EPR1:} a (classical) reality criterion according to which a physical quantity is real if its value can be predicted with certainty without in any way disturbing the system; and

{\bf EPR2:} a criterion of separability (or locality), which was articulated more clearly in subsequent presentations of the argument by Einstein \cite{e11,e12,e13}, according to which ``the real factual situation of system $\mathcal{S}_B$ is independent of what is done with system $\mathcal{S}_A$ if the latter is spatially separated from the former''.
\end{quote}
With these two assumptions EPR showed that the quantum mechanical wave function does not provide a complete description of physical reality.

CQT modifies EPR1 by replacing this ``classical realism'' assumption by ``quantum realism'', whereby physical properties of a system are real \emph{relative} to a well-defined framework, but properties belonging to incompatible frameworks are not \emph {simultaneously real}. As recognized by the authors themselves, the EPR conclusion then does not follow. To quote them ``... one would not arrive at our conclusion if one insisted that two or more physical quantities can be regarded as simultaneous elements of reality \emph{only when they can be simultaneously measured or predicted}'' (italics in the original). Whether one follows EPR or CQT in defining ``physical reality'', the issue in contention concerns the simultaneous reality of quantities pertaining to the \emph{same} subsystem (i.e. $S_x^A$ and $S_z^A$, both of which refer to the spin $S^A$), as opposed to correlations or ``entanglements'' between quantities referring to distant spins $S^A$ and $S^B$. Note also that Bohr's response to EPR \cite{bohr13,bohr14,e11} involved objections to EPR1, phrased in terms of incompatible measurements, and not any denial of EPR2.

In a sense the main criticism leveled by Einstein, in EPR and elsewhere, against orthodox quantum mechanics concerned the notion of collapse of the wave function brought about by a measurement. In the orthodox formulation such a collapse occurs even if the physical action is carried out arbitrarily far from the system being measured. It is the effects associated with collapse, e.g. that $S_z^A$ comes into existence or ceases to exist through a measurement carried out on the arbitrarily distant spin $S^B$, that Einstein termed ``spooky action at a distance'', not the correlations that allow one to determine the value of $S_z^A$ by making a measurement of $S_z^B$. Such distant correlations (within a given framework) occur classically as well as quantum mechanically. According to CQT it is only the mixing of correlations in different quantum frameworks that produces puzzling or incorrect results.

Classical entanglement was described by John Bell using the example of his colleague Dr. Bertlmann (see p. 139 of \textcite{b8}), who always wore socks of different colors. Thus if I see a black sock on Dr. Bertlmann's right foot I know for sure that there is a white sock on his left foot, without having to look. Such entanglements also operate between classical particles, in that I can determine the spin of particle A by measuring the spin of particle B, no matter how far away, provided the two particles were in a state of zero angular momentum at some earlier time and they drifted away without further interactions. An example of distant classical correlations (although not classical entanglements) is the statement that if my daughter has a baby in Sydney, Australia, I instantaneously become a grandfather in NY!

Turning to quantum mechanics, within a given framework entanglement also operates classically. What is special about two quantum mechanically entangled particles is that the correlation is effective in each and every framework, but different correlations that cannot be incorporated into a single framework are not simultaneously meaningful.

As is well known it was \textcite{bell14} who showed that the two EPR assumptions not only lead to a puzzling result, but in fact they contradict the quantitative statistical predictions of quantum mechanics. Indeed, starting from the two EPR assumptions, Bell was able to derive what is known as the Bell inequality, which is violated by quantum correlations. If one believes quantum mechanics, one is thus forced to reject either classical realism, or locality, or both. A number of authors, including Bell himself \cite{bell14,b8}, \textcite{bohm15}, many philosophers and commentators on quantum mechanics \cite{nonl16, nonl17, nonl19} as well as current researchers (see e.g. \textcite{nonl20,nonl21}), favor rejecting locality, but CQT most definitely opts for retaining locality (EPR2) and rejecting classical realism (EPR1), since the latter violates the single-framework rule.

In our view the difference of opinion arises in part because of a confusion regarding the assumptions needed to prove the Bell inequality. Although \textcite{bell14} initially presented the argument as simply drawing a further conclusion from the two EPR assumptions, in later presentations (see \textcite{b8}, p.232) he based the argument entirely on a condition he called ``local causality'', often referred to as ``Bell locality'' \cite{nor1}. Its mathematical expression is
\begin{subequations}
\begin{eqnarray}\label{e11}
\!\text{Prob}(A|B, a,b,\lambda)& = & \text{Prob}(A|a,\lambda), \\
\!\text{Prob}(B|A, a,b,\lambda)& = & \text{Prob}(B|b,\lambda), \\
\!\text{Prob}(A,B|a,b,\lambda)& = &\text{Prob}(A|B,a,b,\lambda)\text{Prob}(B|a,b,\lambda)\nonumber\\
& = &\text{Prob}(A|a,\lambda)\text{Prob}(B|b,\lambda),
\end{eqnarray}
\end{subequations}
where $\text{Prob}(C|D)=\text{Prob}(C,D)/\text{Prob}(D)$ is the conditional probability of $C$ given $D$ and $\text{Prob}(C,D)$ is the joint probability of $C$ and $D$. In Eqs.(11a-c) $A$ and $B$ are measurement outcomes for measurements of types $a$ and $b$, respectively, on systems $\mathcal{S}_A$ and $\mathcal{S}_B$, and $\lambda$ denotes a set of additional variables that may be required to specify \emph{fully} the state of the joint systems. Although it is not completely clear from the discussion in \textcite{b8} referred to above what precise assumptions went into the justification of Eq.(11), Bell and his followers have concluded from the violation of the Bell inequality that follows from (11) that it is the locality assumption that is violated by quantum mechanics. On the other hand many authors have over the years pointed out that locality alone does not justify Eq. (11) or therefore the Bell inequality [see e.g. \textcite{bal1} or more recently \textcite{grif18} and \textcite{tres1}]. In particular the existence of $\lambda$ variables in Eq.(11) that fully specify the state of a physical system depends essentially on some assumption additional to locality, which has been variously named ``predictive completeness'', ``classical (or weak) realism'', or ``counterfactual definiteness'', as well as other designations. According to these authors (and to CQT) it is this additional assumption that is contradicted by quantum mechanics, and referring to Eq.(11) simply as a locality assumption is at best misleading. Indeed, the violation of Eq. (11) by quantum mechanics is due to the incompatibility between different local choices for $a$, i.e. $S_x^A$ or $S_z^A$ (or different choices for $b$), rather than any ``spooky interaction'' between the distant spins $S^A$ and $S^B$, so it is unrelated to the locality property EPR2. In particular, \textcite{grif18} argues that there are no quantum variables $\lambda$ that satisfy Eq.(11) if the different choices for $a$ (or for $b$) do not commute. What is at issue here is not the fact that Eq.(11) and the Bell inequality that follows from it are violated by quantum mechanics. The issue is to understand the origin of the violation and whether it is associated with a failure of realism or completenes (EPR1), or of separability (EPR2). Needless to say, there is no consistent framework for which Eq.(11) is satisfied.

The present author believes that had Einstein been aware of Bell's result, which follows from EPR1 and EPR2, he too would have retained locality, since it is so closely tied to the requirements of relativity, and he would have had to reject classical realism. It is perhaps permitted to fantasize that Einstein would have embraced the quantum realism of CQT, as the best way to reconcile quantum mechanics (whose predictions Einstein did not doubt) with the requirements of locality. According to CQT reality is \emph{relative} to a framework just as simultaneity is relative to a reference frame in special relativity.

\section{\label{8} Consistent histories}
In order to provide a complete formulation of quantum mechanics the foregoing notions must be generalized to consider systems with arbitrary numbers of degrees of freedom as well as their time dependence. This generalization is the CQT formulated by \textcite{grif2} and elaborated by others \cite{gh3,o4}, as described in detail in the book by \textcite{grif1}. The general theory is still based on the fundamental principle that what characterizes any quantum system is the existence of multiple incompatible frameworks, which are a generalization of the frameworks defined above.

Let us first define a ``history'' as a set of ``events'' or ``properties'', represented by projection operators on the Hilbert space at a sequence of times $t_1,... t_n$. We then consider a collection, or ``family'' of histories as a candidate for a (consistent) framework. The probability of a given history is calculated using a generalization of the Born rule, as well as the dynamical properties of the system, whose states evolve according to the Schroedinger equation governed by the full Hamiltonian of the system (see below). In contrast to the spin-\mbox{\textonehalf} case at a single time, where frameworks were more or less automatically consistent, it is in general necessary to impose stringent \emph {consistency} conditions in order for a given family of histories to provide valid quantum predictions. The conditions for a family to constitute a consistent framework are:
\begin{quote}
(i)	The sum of probabilities of the histories in the family is unity;

(ii)Two distinct histories are mutually orthogonal, i.e. their product vanishes.
\end{quote}
Each framework can again be described as an Exhaustive Set of Exclusive Alternatives (ESEA), exhaustive because of (i) and exclusive because of (ii), with a suitable definition of the product. Also, just as for spin-\mbox{\textonehalf}, frameworks cannot be combined, i.e. you obtain incorrect predictions if you take a subset of histories from one framework (with probabilities adding up to $p$, say) and combine them with another subset of histories from another framework (with probabilities $1-p$), in violation of the single-framework rule. We shall use the terms family and framework somewhat interchangeably, but in general reserve the latter term for families whose histories satisfy the consistency conditions. Another term for frameworks is ``realms'' \cite{gh3}.

Specifically, let $t_0,t_1,...,t_n$ be a discrete set of times and let a ``property'' or ``event'' $\alpha_j$ at time $t_j$ be defined as a state or set of states in the Hilbert space of the system under study, or as the projection operator $P(\alpha_j)$ onto that state or set of states, for which we shall use the same notation. A ``history'' is then defined as a sequence of projection operators whose tensor product is
\begin{eqnarray}\label{e12}
C(\alpha)&=&C(\alpha_0,\alpha_1,...,\alpha_n)\nonumber\\
         &=& P(\alpha_0)\odot P(\alpha_1)\odot...\odot P(\alpha_n),
\end{eqnarray}
where the notation $\odot$ is designed to indicate that each $P(\alpha_k)$ acts in a copy of the system Hilbert space at time $t_k$, so that $C(\alpha)$ is an operator in the ``history Hilbert space'' consisting of $n$ replicas  of the original Hilbert space. In general the product of two histories must be defined as an operation in this history Hilbert space (see Ch.11 of \textcite{grif1}).
If the system under study is assumed to be in the pure state $|\psi_0\rangle$ at time $t_0$, however, i.e. if the projector $P(\alpha_0)$ is given by
\begin{equation}\label{12.5}
P(\alpha_0)=|\psi_0\rangle \langle\psi_0| \equiv [\psi_0],
\end{equation}
then it is possible to define the product of two histories in terms of ``chain kets'' in the original Hilbert space as
\begin{equation}\label{12.6}
\langle C(\alpha),C(\beta)\rangle = \langle \alpha|\beta\rangle,
\end{equation}
where
\begin{equation}\label{12.7}
|\alpha\rangle = \hat{P}(\alpha_n)...\hat{P}(\alpha_1)|\psi_0\rangle,
\end{equation}
and $\hat{P}(\alpha_k)$ is a projection operator at time $t_k$ in the Heisenberg representation. The probability of the history  $C(\alpha)$ is then given by a generalization of the Born rule
\begin{equation}\label{e13}
\text{Prob}[C(\alpha)]=\langle C(\alpha),C(\alpha)\rangle = \langle\alpha|\alpha\rangle.
\end{equation}
Let us now consider a set or family of histories
\begin{equation}\label{e14}
\mathcal{F}(\{\alpha\})=\{C(\alpha^{(1)}),C(\alpha^{(2)}),...C(\alpha^{(m)})\},
\end{equation}
where each of the $C(\alpha^{(k)})$ is a history, i.e. a product of $n$ projection operators as in Eq.(\ref{e12}). In order to be able to assign probabilities in a consistent manner, the family $\mathcal{F}(\{\alpha\})$ must satisfy the following conditions:
\begin{equation}\label{e15}
\sum_{k=1}^{m} C(\alpha^{(k)})=I_{ij} ,
\end{equation}
where $I_{ij}$ is the $n\times n$ unit matrix in the history Hilbert space and
\begin{equation}\label{e16}
\langle C(\alpha^{(k)}), C(\alpha^{(k^\prime)})\rangle =0,\; \text{for}\; k \neq k^\prime,
\end{equation}
which expresses the orthogonality condition (ii).
A family of histories satisfying Eqs.(\ref{e15}) and (\ref{e16}) is called a ``consistent framework'' (or framework for short). Equation(\ref{e15}) ensures that the framework is exhaustive and Eq.(\ref{e16}) implies that the different histories are exclusive, i.e. they do not ``interfere'', thus permitting an interpretation in terms of standard (Boolean) probabilities within each framework.

The hallmark of a quantum, as opposed to a classical, system is the existence of more than one (in fact an infinite number of) framework(s). Any statement about a quantum system is expressed in the form
\begin{equation}\label{e17}
\text{Prob}[C(\alpha^{(k)})]=p_k,
\end{equation}
i.e. the probability of the $k^{th}$ history in the framework $\mathcal{F}(\{\alpha\})$ is $p_k$. Such statements refer to ``real'' histories of the form (\ref{e12}), which consist of sequences of ``real'' properties, represented by projection operators onto states $|\alpha_j\rangle$ of the system. The reality of these properties, however, is \emph{relative} to a particular framework.

For illustration let us first consider once more the free spin-\mbox{\textonehalf} discussed above. We denote the projector onto the state $|w^+ \rangle$ at time $t_j$ by
\begin{equation}\label{e18}
P(|w^+ \rangle,t_j)=[w_j^+]=|w^+(t_j) \rangle \langle w^+(t_j)|.
\end{equation}
A history with times $t_0, t_1,...t_n$ will be written as
\begin{equation}\label{e19}
C=[u_0]\odot [v_1]\odot ... \odot  [w_n],
\end{equation}
in the history Hilbert space, where $[u_i], [v_j], [w_k]$ denote projection operators onto states associated with various directions in 3-space at times $t_i, t_j, t_k$, respectively,  e.g. $[u_i]=P(|u^+ \rangle,t_i)$, using the same notation as in Eq.(\ref{12.5}) above.

It can be shown that any exhaustive family of two-time histories automatically satisfies the consistency conditions (\ref{e15}) and (\ref{e16}). For three-time histories, on the other hand, the consistency conditions are not necessarily satisfied. An example is the family consisting of the histories
\begin{subequations}
\begin{eqnarray}\label{e20}
C^{(1)} & = & [z_0^+] \odot [x_1^+] \odot [z_2^+],\\
C^{(2)} & = & [z_0^+] \odot [x_1^+] \odot [z_2^-],\\
C^{(3)} & = & [z_0^+] \odot [x_1^-] \odot [z_2^+],\\
C^{(4)} & = & [z_0^+] \odot [x_1^-] \odot [z_2^-] .
\end{eqnarray}
\end{subequations}
The framework consisting of the histories in (23) can be shown to be inconsistent since the histories $C^{(1)}$ and $C^{(3)}$ [or $C^{(2)}$ and $C^{(4)}$] are not mutually orthogonal, i.e. they interfere. It is the presence of the final $[z_2^+]$ operator in Eqs.(23a) and (23c) that brings about the nonorthogonality or interference between the histories $C^{(1)}$ and $C^{(3)}$. If $[z_2^+]$ is replaced by a unit operator, on the other hand, then the histories $C^{(1)}$ and $C^{(3)}$ become orthogonal. Another way to make the framework (23) consistent is to apply an appropriate magnetic field during the time interval $t_0<t<t_2$, in such a way that the history $C^{(2)}$ becomes unitary and the histories  $C^{(1)}, C^{(3)}$ and $C^{(4)}$ vanish. Alternatively, if we consider a free spin subject to a measurement of $S_x$ at $t=t_1$, followed by a measurement of $S_z$ at $t=t_2$, then the physical acts of measurement allow one to define consistent histories with the expected probabilities for experimental outcomes \cite{grif1}. Thus the consistency of a framework is not a matter of arbitrary choice or convention but instead it depends on the forces and constraints applied to the system under study.

Let us now consider briefly two entangled spins $S^A$ and $S^B$, represented by operators $[z^+_{Aj}], [z^+_{Bj}]$, etc. at times $t_j$. Let the initial state $|\psi_0\rangle$ be the singlet
\begin{equation}\label{e21}
|\psi_0\rangle = (1/\sqrt{2})(|z^+_{A0}\rangle |z^-_{B0}\rangle-|z^-_{A0}\rangle |z^+_{B0}\rangle).
\end{equation}

Then we may define the two-time family consisting of the histories
\begin{subequations}
\begin{eqnarray}\label{e22}
C^{(1)} & = & [\psi_0] \odot [w_{A1}^+] [v_{B1}^+],\\
C^{(2)} & = & [\psi_0] \odot [w_{A1}^-] [v_{B1}^+],\\
C^{(3)} & = & [\psi_0] \odot [w_{A1}^+] [v_{B1}^-],\\
C^{(4)} & = & [\psi_0] \odot [w_{A1}^-] [v_{B1}^-],
\end{eqnarray}
\end{subequations}
which may be shown to be consistent for arbitrary directions $\bf{v}$ and $\bf{w}$.

Turning now to three-time histories we can consider the consistent framework with a single history
\begin{equation}
C= [\psi_0]\odot [\psi_1]\odot [\psi_2],\label{e23}
\end{equation}
whose probability is unity, where $[\psi_j]$ is the projector onto the state $|\psi_0 \rangle$ advanced by the unitary Schroedinger time dependence to the time $t_j$. We refer to the family (\ref{e23}) as the ``unitary framework'' (see Sec. IX below). Another consistent framework splits the histories at $t=t_1$ and then continues unitarily to $t=t_2$
\begin{subequations}
\begin{eqnarray}\label{e24}
C^{(1)} & = & [\psi_0] \odot [z_{A1}^+] [z_{B1}^-]\odot [z_{A2}^+] [z_{B2}^-],\\
C^{(2)} & = & [\psi_0] \odot [z_{A1}^-] [z_{B1}^+]\odot [z_{A2}^-] [z_{B2}^+],
\end{eqnarray}
\end{subequations}
to yield two histories each with probability \mbox{\textonehalf}. Alternatively, the split could occur at $t=t_2$, and by spherical symmetry of $|\psi_0 \rangle$ the direction $z$ in Eqs.(27) can be replaced by any other direction $\bf{w}$.
As an example of an inconsistent family we show the set of 16 histories
\begin{eqnarray}
[\psi_0] &\odot &\{[x_{A1}^+], [x_{A1}^-]\}\{[z_{B1}^+], [z_{B1}^-]\}\nonumber\\
&\odot& \{[z_{A2}^+], [z_{A2}^-]\}\{[x_{B2}^+], [x_{B2}^-]\},\label{e25}
\end{eqnarray}
where the notation $\{C,D\}\{E,F\}$ denotes the four products $CE,DE,CF,DF$. It can then be verified that the family (28) is inconsistent since there are four histories ending in $[z_{A2}^-] [x_{B2}^- ]$ and these are not mutually orthogonal.

Finally, it should be noted for clarity that the histories of CQT are different from the histories in Feynman's path integral (sum over histories) formulation of the solution of the Schroedinger equation \cite{fh19}. In the Feynman case the histories are associated with complex amplitudes so they interfere with one another, whereas the histories of CQT are mutually exclusive and only appear in terms of their (real) probabilities in any physical predictions. The mysterious physics of quantum superpositions is fully captured in CQT by the existence of multiple incompatible frameworks, each one of which is in a sense classical (Boolean), not by interference of histories or any other 'superposition'.

The above brief description of CQT is intended as an informal summary of its main qualitative features. A more detailed description can be found in the book by \textcite{grif1} and references therein. In particular Ch. 23 provides a detailed illustration of the different frameworks associated with singlet state correlations of two spins.
\section{\label{9} The unitary framework and the collapse framework}

In the language of CQT one can define the so-called unitary framework, which in one guise or another is implicit in many discussions of quantum mechanics. We consider an arbitrary system with initial state $|\psi_0 \rangle$ at $t=t_0$, and let it evolve deterministically according to the Schroedinger equation with some given Hamiltonian. We denote by $|\psi_n \rangle$ this exact state at time $t=t_n$. Then the unitary framework consists of the single history
\begin{equation}\label{e27}
[\psi_0] \odot [\psi_1] \odot ...\odot [\psi_n],
\end{equation}
whose probability is unity. This is a valid prediction of quantum mechanics, but unless the final state is the eigenstate of some straightforward observable it is also not a particularly useful prediction, since it contains no accessible physical information. In particular the values of any observables are in general undefined in this framework.

There is, however, a framework that is close to the unitary one, which we can call the ``collapse framework'', which is the one employed in the orthodox formulation of quantum mechanics. It consists of the histories enumerated by the index $k$,
\begin{equation}\label{e28}
C_k^A = [\psi_0] \odot [\psi_1] \odot ...\odot [\psi_{n-1}] \odot [A_{kn}],
\end{equation}
where $[A_{kn}]$ is the projector onto the $k^{th}$ eigenfunction of the operator $A$ at time $t_n$, with eigenvalue $a_k$ and $A$ is some observable. Since the eigenfunctions of $A$ form an orthonormal basis, the family $C_k^A$ is an appropriate consistent framework. If the Hamiltonian of the system is such that a measurement of $A$ is carried out at the time $t_n$, then this framework embodies the orthodox notion that in all histories the system is described by $|\psi(t) \rangle$  from $t=t_0$ to $t=t_{n-1}$, and that the system then ``divides'' into the different eigenfunctions of $A$ when a measurement of $A$ is made at $t=t_n$. In CQT this collapse framework is one valid representation of the system, but by no means the only one and it does not signify that any physical collapse has taken place.
\section{\label{10}Quantum paradoxes}
CQT, as the name indicates, is designed to provide a consistent account of the predictions of quantum mechanics, grounded in a realistic view of the microscopic world, that recognizes the profoundly counter-intuitive nature of complementarity but defines its consequences precisely. The theory thus contains no paradoxes, so it must in some way ``explain'' the well-known quantum paradoxes that have been the subject of so much discussion in the literature. Generally speaking, each paradox can be seen, from the point of view of CQT, as a more or less subtle violation of the \emph {single-framework rule}, as explained in detail in Chapters 20-25 of the book by \textcite{grif1}. We have already indicated how this occurs for the EPR and Bell ``paradoxes'', and we shall confine our further remarks to a brief discussion of the Schroedinger Cat paradox, since it has been cited by \textcite{leg20,leg21} as the most important problem with orthodox quantum mechanics.

The Schroedinger Cat paradox can be thought to arise because one is implicitly thinking in terms of the unitary framework, and at the same time thinking of the system as containing a superposition of live and dead cats. According to CQT, however, there are no cats at all in the unitary framework and in the cat-framework there are no superpositions. The simple analogue for the spin-\mbox{\textonehalf} system in the state $|z^+ \rangle$ is to say ``I will accept a superposition of $|x^+ \rangle$  and $|x^- \rangle$ for microscopic states, since we know from experience that the microworld is weird, but for macroscopic (cat) states, that is impossible''. The answer from CQT is that in neither case (micro or macro) is the ``superposition'' a valid quantum mechanical description. In the z-framework neither $|x^+ \rangle$ nor $|x^- \rangle$  is defined (no cats) and in the x-framework we have either $\langle S_x \rangle =\mbox{\textonehalf}$ (live cat) or $\langle S_x \rangle =-\mbox{\textonehalf}$ (dead cat), each with probability \mbox{\textonehalf}. For the macroscopic case the analogue of the z-framework is the unitary framework and it has no simple physical interpretation, and certainly no cats. The analogue of the x-framework is the set of (Boolean) alternatives -live cat or dead cat -, with appropriate probabilities for each.

\section{\label{11}Decoherence and the classical limit}

The effects of a random environment on physical systems and measurement instruments, termed decoherence \cite{z21,z9}, are often cited as explaining how one passes from the microscopic to the macroscopic and thus how the collapse invoked in the orthodox formulation of quantum mechanics is to be interpreted. Decoherence is also invoked to understand the ``measurement problem'' and the classical limit of quantum mechanics. From this point of view there is nothing special about measurements: they are a particular type of physical interaction that needs to be explained by the theory, rather than being an irreducible part of its formulation as in the orthodox or Copenhagen viewpoints.

It has been pointed out, however, that although decoherence does make plausible the passage from quantum to classical domains, when it is applied to the full quantum mechanical wave function as a representation of the state of the system, it still does not by itself remove the measurement problem or the paradox of Schroedinger's Cat \cite{leg20,adler22}.

CQT agrees with the above statements since, as stated previously, it attaches no special significance to macroscopic measurements. Moreover, the theory recognizes that decoherence applied in the unitary framework does indeed not resolve the paradox of Schroedinger's Cat or the measurement problem. The relevant question, however, is ``what is the effect of decoherence not on wavefunctions but on \emph{histories}, in the \emph{different} frameworks?'' Although the answer has not been fully worked out, the problem has been extensively studied, primarily by \textcite{gh3}, and by \textcite{o4} and the overall picture is the following (see also Ch 26 of \textcite{grif1}).

Decoherence in general destroys interference effects and it can thus render certain families of histories (frameworks) consistent, when in the absence of decoherence they would be inconsistent. It can then be argued, and shown in detail for simple models, that a large class of ensuing coarse-grained histories (as opposed to wavefunctions), are close to the corresponding classical trajectories for the system and that the corresponding probabilities are close to unity, thus justifying the determinism of classical mechanics as \emph{emerging} from quantum mechanics. This does not mean that all frameworks are transformed into sets of quasiclassical orbits, for example the unitary framework certainly is not, but the latter also does not in general provide a useful set of predictions. The general situation appears to be that in the classical limit, frameworks are either inconsistent (thus meaningless), uninformative (e.g. the unitary framework), or they approach classical behavior.

In this way CQT naturally leads to the emergence of quasiclassical frameworks on the coarse-grained scale of macroscopic phenomena, coexisting with a multiplicity of incompatible but individually consistent frameworks on the microscopic scale. As noted, such a division between the classical and quantum domains was a starting assumption of the orthodox and Copenhagen formulations. Note that decoherent histories are a special, albeit important, example of consistent histories, but the physical phenomenon of decoherence is not necessary in order to formulate the theory, for example it is not involved in our description of the single spin-$\mbox{\textonehalf}$ system in Sec. II, so Griffiths's consistency conditions represent a more general formulation of the approach.

\section{\label{12}Concluding remarks}
It appears that the main philosophical difference between CQT and most other formulations of quantum mechanics lies in the role ascribed to the wavefunction. Since the wavefunction encodes all the information about a system it is natural to interpret it to ``represent'' the system, just as the phase space point and its orbit represent a classical system. It was recognized early on by Bohr and his colleagues, however, that the wavefunction could not provide a \emph{realistic} representation of a quantum system, isolated from observation, i.e. from some measurement apparatus. This idea was made precise by EPR, who concluded that the wavefunction does not provide a ``complete description'' of a quantum system. \textcite{bohm15} and \textcite{bell14} then showed that if one insists on a (classically) realistic representation, it follows that quantum systems will inevitably display nonlocal influences that violate the spirit, if not the letter, of relativity.

In its most extreme form the insistence that the wave function provide a complete representation of quantum reality leads to the ``relative state'' or ``many worlds'' theory of \textcite{ev1}. The dizzying complexity of even the simplest systems in this theory seems to the present author to provide ample motivation for seeking a different formulation.

CQT, on the other hand, retains the requirement of realism, but it abandons the notion that it is the wave function that ``represents'' the system under study, substituting instead histories, provided these are part of a consistent framework. All of the difficulties encountered by previous formulations and interpretations (the paradoxes and the weirdness) are encapsulated (not eliminated!) in the notion that quantum systems necessarily possess \emph{multiple incompatible frameworks}. Once this notion has been accepted (``digested'' might be a better term) the theory unfolds in a logical manner without further paradoxes or mysteries. In CQT the wavefunction is by no means unimportant, since as mentioned above it encodes all the dynamical information about a system and it is thus essential for calculating any meaningful properties, but it does not provide a direct universal representation of quantum reality. It is true, of course, that in the unitary framework described in Sec. IX the wave function constitutes a valid history, but that history is in general not useful for answering physical questions about the system.

In a sense, CQT can be considered to vindicate EPR in that it confirms the statement that the wavefunction does not provide a complete representation (``description'') of physical reality. CQT also follows EPR in seeking a realistic formulation of quantum mechanics, but it replaces the classical realism of EPR1 by the quantum realism of multiple coexisting but incompatible frameworks. In that sense CQT reconciles Einstein and Bohr, but without the latter's resort to measurement and classical apparatus in the definition of quantum reality.

It may also be argued that CQT satisfies the requirements put forth by \textcite{b7} in his criticism of the orthodox and Copenhagen viewpoints. CQT is not based on measurement, it does not require an artificial separation (a ``cut'') between the classical and quantum domains and it needs no ``interpretation'', since the theory is defined by the equations themselves.

We have already discussed the sense in which the violation of the Bell inequality, which results from the violation of Eq.(11), should be interpreted as negating classical realism rather than locality. As emphasized by \textcite{grif18}, quantum mechanics certainly has nonlocal states and frameworks, but then so does classical mechanics, for example a plane wave state is surely nonlocal. What does not occur in quantum mechanics, any more than in classical mechanics, is true nonlocality in the sense of nonlocal \emph{influence} of one localized subsystem on another remote localized subsystem. This is why the pervasive use of the term ``nonlocal'' to describe quantum mechanics and particularly the violation of Eq.(11), seems to us to be maximally misleading.

Let us comment briefly on the concept of ``entanglement'', which is cited by many authors as the signature property of quantum systems that illustrates its fundamental nonlocal character. From the point of view of CQT, entanglement is a description of how properties contained in framework $\mathcal{F}$ appear in the language of framework $\mathcal{G}$. The phenomenon is particularly striking when properties referring to two widely separated subsystems factorize in framework $\mathcal{F}$, while those same two subsystems contribute to a single property in framework $\mathcal{G}$. An example is the EPR pair considered in Eq.(\ref{e21}) above, for which one can calculate the probability that spins A and B have a relative angle $\theta$ using the framework (25), whereas these properties are fully entangled and thus undefined in the unitary framework based on (\ref{e21}). Similarly, in a two-slit experiment the framework displaying the interference pattern does not contain states in which the particle goes through either slit, since these are entangled in that framework. Strictly speaking, CQT does away with entanglement and replaces it with incompatible frameworks, i.e. the knowledge that even though a framework is exhaustive it is not ``complete'', since it allows for the existence of other frameworks.

Quantum reality can be succinctly defined as ``any property or set of properties of a system for which quantum mechanics can calculate a probability''. Note that quantum mechanics does not answer the question ``which real property or event actually occurs?'' That question can be considered ``excess baggage'' \cite {jh2}. A further essential insight of CQT is that probabilities can only be defined if a suitable sample space has been specified, and quantum systems are characterized by a multiplicity of overlapping and incompatible sample spaces, called frameworks. Any statement about quantum reality must specify the framework (or set of frameworks) to which it applies.

We conclude by quoting Richard Feynman \cite{feyn1}:
\begin{quote}
In this chapter we shall tackle immediately the basic element of the mysterious behavior in its most strange form. We choose to examine a phenomenon [two-slit interference] which is impossible, \emph{absolutely} impossible, to explain in any classical way, and which has in it the heart of quantum mechanics. In reality, it contains the \emph{only} mystery. We cannot make the mystery go away by ``explaining'' how it works. We will just \emph{tell} you how it works. In telling you how it works we will have told you about the basic peculiarities of all quantum mechanics.
\end{quote}

Feynman's point of view is very much in line with that of CQT, in particular the statement that the \emph{only} mystery of quantum mechanics can be fully displayed using the example of a single electron in a two-slit interference experiment, which displays the wave-particle duality. Note in particular the absence of any reference to entanglement of distant particles or to nonlocality of quantum behavior.
\begin{acknowledgments}
The author wishes to thank Robert Griffiths, James Hartle, Tony Leggett, David Mermin and especially Richard Friedberg for useful discussions and comments on the manuscript. He also acknowledges the hospitality of the Aspen Center for Physics where this paper was largely written.
\end{acknowledgments}

\bibliographystyle{apsrmp}
\bibliography{cqt-ref}

\newpage

\end{document}